\documentstyle[12pt,epsfig,aps]{revtex}

\begin{document}
   
\author{ U\v{g}ur TIRNAKLI\thanks{e-mail: tirnakli@fenfak.ege.edu.tr}$\;$ , 
Fevzi B\"{U}Y\"{U}KKILI\c{C} , Do\v{g}an DEM\.{I}RHAN \\
Department of Physics, Faculty of Science,\\ Ege University 
35100, Bornova \.{I}zmir-TURKEY}

\title{Generalized Distribution Functions and an Alternative Approach to
Generalized Planck Radiation Law }

\date{\today}

\maketitle
   
\begin{abstract}

In this study, recently introduced generalized distribution functions are
summarized and by using one of these distribution functions, namely
generalized Planck distribution, an alternative approach to the generalized
Planck law for the blackbody radiation has been tackled. The expression
obtained is compared with the expression given by C. Tsallis et al. [Phys.
Rev. E52, (1995) 1447], and it is found that this approximate scheme
provides bounds to the exact result, depending on the values of $q$ index.

\end{abstract}

\section{Introduction}

Although the necessity of nonextensive statistical thermodynamics has been
very clear for a long time in many physical systems such as
three-dimensional self-gravitating astrophysical objects [1], black holes
and superstrings [2], L\'{e}vy-type random walks [3], vortex problem [4] and
dark matter [5] where long-range microscopic interactions are present; an
increasing tendency towards nonextensive formalisms keeps growing nowadays,
along two lines: Quantum-Group-like approaches and Generalized Statistical
Thermodynamics (GST).

GST is recently introduced by C. Tsallis [6,7] and then the formalism not
only has been applied to numerous concepts of statistical thermodynamics
[8-28], but also prosperous for the physical systems [29-33] where extensive
Boltzmann-Gibbs statistics fails. The detailed reviews of the general
aspects of the formalism can be found in [34], and some investigations of
the subject from the mathematical point of view are now available in [35].

The formalism has been based upon two axioms; namely (i) the entropy of the
system is given by [6]

\begin{equation}
S_q=-k\frac{1-\sum_{i=1}^{W}p_i^q}{1-q}
\end{equation}

\noindent where $k$ is a conventional positive constant, $p_i$ is the
probability of the system to be in a microstate, $W$ is the total number of
configurations of the system and $q$ is a new parameter which is especially
called the Tsallis $q$-index; (ii) the $q$-expectation value of an
observable $\cal{O}$ is given by 

\begin{equation}
\left<{\cal O}\right>_q=\sum_{i=1}^{W}p_i^q {\cal O}_{i} \;\;\;\; .
\end{equation}

\noindent On the other hand, GST contains Boltzmann-Gibbs Statistics as a
special case when $q=1$. Clearly $S_q$ is extensive if and only if $q=1$ and
$(1-q)$ can be interpreted as the measure of the lack of extensivity of the
system [24].

\section{Generalized Distribution Functions}

Amongst the systems handled in the frame of GST, the classical and quantum
gases have been investigated for the first time by two of the authors of
the present study and the results have been presented in two seperate
manuscripts [13,14]. In the first paper, the fractal inspired entropies of
the classical and quantum gases have been found by making use of the
statistical weights and the generalized distribution functions have been
obtained by introducing the constraints related to the statistical
properties of the particles into the entropy [13]. In the second paper, the
Tsallis entropy have been used for the same purpose and the same conclusions
for the generalized distribution functions have been attained but in this
case the constraints concerned with the statistical properties of the
particles have been introduced in the calculation of the partition
functions. However, the factorization of the partition function which was
used in this calculations has been the main disputable point of the
approximation. We claim that this approximate procedure provides a bound to
the exact results, depending on the values of $q$. In order to prove this,
let us take a very simple physical system which could be in bounded or free
states with the energy levels $A$ and $B$. For this system let us take the
inequality 

\begin{equation}
\left[1+(1-q)(A+B)\right]^{\frac{1}{1-q}}\neq\left[1+(1-q)A\right]^
{\frac{1}{1-q}}
\left[1+(1-q)B\right]^{\frac{1}{1-q}}\;\; .
\end{equation}

\noindent It is easy to rewrite this in exponential form

\begin{eqnarray}
\exp\left\{\frac{1}{1-q}\ln\left[1+(1-q)(A+B)\right]\right\}\neq &
\nonumber \\
\exp\left\{\frac{1}{1-q}\ln\left[1+(1-q)A\right]+
\frac{1}{1-q}\ln\left[1+(1-q)B\right]\right\}\; .
\end{eqnarray}

\noindent After some algebra it is straightforward to find

\begin{eqnarray}
\exp\left\{\frac{1}{1-q}\ln\left[1+(1-q)A+(1-q)B\right]\right\}\neq &
\nonumber \\ 
\exp\left\{\frac{1}{1-q}\ln\left[1+(1-q)A+(1-q)B+(1-q)^{2}AB\right]\right\}
\end{eqnarray}

\noindent If the left-hand side and right-hand side of this inequality are
called "exact" and "approximation" respectively, then it follows that
exact$>$approximation for $q>1$ whereas exact$<$approximation for $q<1$. It
must be emphasized that this situation is valid if and only if the product
$A.B$ is positive. In photon case, which is the subject of this study, $A$
and $B$, namely energy levels of the system are defined to be $h\nu /kT$
that is always positive.

In addition to this, an inequality can be derived for the partition function
within the frame of (3). The grand partition function reads :

\begin{equation}
Z_q=\sum_{\{r_i\}}\left[1-\beta (1-q)\sum_{r=1}^{m}\left(\epsilon
_r-\mu\right)\right]^{\frac{1}{1-q}}\;\;\; ,
\end{equation}

\noindent where $\{r_i\}$ are the elements of the grand canonical ensemble.
For the photon case $(\mu =0)$, the eq.(6) becomes

\begin{equation}
Z_q=\sum_{\{r_i\}}\left[1-\beta (1-q){\cal
H}_{\{r_i\}}\right]^{\frac{1}{1-q}}
\end{equation}

\noindent where ${\cal H}_{\{r_i\}}=\sum_{r=1}^{m} \epsilon _r$ is the
Hamiltonian of the system. Employing the inequality (3), another inequality
for the partition function can be written down as

\begin{eqnarray}
\sum_{\{r_i\}}\left[1-\beta (1-q)\left(\epsilon _1+\cdots +\epsilon
_m\right)\right]^{\frac{1}{1-q}}\neq & 
\nonumber \\
\sum_{\{r_i\}}\left[1-\beta (1-q)\epsilon
_1\right]^{\frac{1}{1-q}}\;\cdots\;
\left[1-\beta (1-q)\epsilon _m\right]^{\frac{1}{1-q}}\;\;\; .
\end{eqnarray}

\noindent Since the free energy is a monotonic function of $Z_q$, namely,

\begin{equation}
F_q=-\frac{1}{\beta}\frac{Z_q^{1-q}-1}{1-q}\;\;\; ,
\end{equation}

\noindent it is straightforward to write an inequality for the free energy.

In the frame of the above simple mathematical analysis, it is evident that
our approximation scheme provides a lower or upper bound to the exact
results, depending on the $q$ values.

Within this approximation procedure the generalized distribution functions
are given by

\begin{equation}
\left<n_r\right>_q^{MB}=\left[1-(1-q)\beta (\epsilon _r -\mu
)\right]^{\frac{1}{1-q}}
\end{equation}

\begin{equation}
\left<n_r\right>_q^{FD}=\frac{1}{\left[1-(1-q)\beta (\epsilon _r -\mu
)\right]^{\frac{1}{q-1}}+1}
\end{equation}

\begin{equation}
\left<n_r\right>_q^{BE}=\frac{1}{\left[1-(1-q)\beta (\epsilon _r -\mu
)\right]^{\frac{1}{q-1}}-1}
\end{equation}

\noindent where $MB$, $FD$ and $BE$ stand for Maxwell-Boltzmann, Fermi-Dirac
and Bose-Einstein, respectively. In a very recent effort [36], quantum
statistics has been studied by means of a kinetic approach and the
distribution functions corresponding to the Tsallis probability are found to
be the same with the eqs.(10)-(12). It is easy to show that the standard
distribution functions are recovered in the $q\rightarrow 1$ limit. On the
other hand, when $\mu$ is set equal to zero in eq.(12), one can obtain the
generalized Planck distribution ($PD$)

\begin{equation}
\left<n_r\right>_q^{PD}=\frac{1}{\left[1-(1-q)\beta\epsilon _r\right]^
{\frac{1}{q-1}}-1}
\end{equation} 

\noindent which is going to be used here for making an approximation to the
derivation of the generalized Planck radiation law by following the
well-known procedure available in many textbooks [37] of Statistical
Physics.

\section{The Generalization of the Planck Radiation Law}

Very recently, the generalization of the Planck radiation law is obtained by
Tsallis et al. [24] in order to see whether present cosmic background
radiation is (slightly) different from Planck radiation law due to
long-range gravitational influence. In the course of their investigations,
making use of the partition function given by
$Z_q\approx Z_1 \left\{1-\frac{1}{2}(1-q)\beta^2 \left<{\cal
H}^2\right>_1\right\}$ (where $Z_1$ and $\left<{\cal H}^2\right>_1$ stand
for the values of these quantities in standard Boltzmann-Gibbs statistics)
in the $\beta (1-q)\rightarrow 0$ limit, Tsallis et al. have achieved the
generalization of the Planck law given in the following expression for
$q\cong 1$ [24]

\begin{equation}
\frac{D_q(\nu)h^2c^3}{8\pi
(kT)^3}\approx\frac{x^3}{e^x-1}\left(1-e^{-x}\right)^{q-1}\left\{1+(1-q)x
\left[\frac{1+e^{-x}}{1-e^{-x}}-\frac{x}{2}\frac{1+3e^{-x}}{\left(1-e^{-x}
\right)^2}\right]\right\}
\end{equation}

\noindent where $D_q(\nu)$ is the photon energy density per unit volume,
$\nu$ is the photon frequency and $x\equiv h\nu /kT$. Then they applied this
expression to the cosmic microwave background radiation to test for
deviations from Planck radiation law and found a 95\% confidence limit of
$|q-1|<3.6\times 10^{-5}$ from the data obtained via Cosmic Background
Explorer Satellite by Mather et al. [38]. In the present work we're trying
to generalize the Planck law by using an approximate scheme which seems
simpler and more general (not necessarily $q\cong 1$).

The distribution of photons among the various quantum states with definite
values of the momentum $h\vec{k}/2\pi$ and energies $h\nu$ can be given by
generalized Planck distribution (eq.(13))

\begin{equation}
\left<n_r\right>_q^{PD}=\frac{1}{\left[1-(1-q)\frac{h\nu}{kT}\right]^
{\frac{1}{q-1}}-1}\;\;\; .
\end{equation} 

\noindent On the other hand, the number of quantum states of photons with
frequencies between $\nu$ and $\nu +d\nu$ is $8\pi V\nu ^2 d\nu /c^3$ ($V$
being the volume of the photon gas). It is clear that multiplying this
quantity by eq.(15), the number of photons in this frequency interval can be
determined:

\begin{equation}
dN=\frac{8\pi V}{c^3}\frac{\nu ^2 d\nu}{\left[1-(1-q)\frac{h\nu}{kT}\right]^
{\frac{1}{q-1}}-1}\;\; .
\end{equation}

\noindent Therefore the photon energy in this interval is given by

\begin{equation}
dE=\frac{8\pi hV}{c^3}\frac{\nu ^3 d\nu}{\left[1-(1-q)\frac{h\nu}{kT}\right]^
{\frac{1}{q-1}}-1}\;\; ,
\end{equation}

\noindent and finally the photon energy density per unit volume is

\begin{equation}
D_q(\nu )=\frac{8\pi h}{c^3}\frac{\nu ^3}{\left[1-(1-q)\frac{h\nu}{kT}\right]^
{\frac{1}{q-1}}-1}\;\; ,
\end{equation}

\noindent which generalizes the Planck's law. The comparison of this
expression with that given by Tsallis et al. (eq.(5) of their paper) is
illustrated in Figure. It is observed in the figure that, at low frequencies
the two $D_q(\nu)$ plots completely fit into one another. On the other hand,
towards the frequency values where the maxima of the curves occur, the plots
diverge from one another by a certain amount. This is an expected result
since the implication of our approximation scheme shows itself as an upper
(lower) bound to the exact result when $q>1$ ($q<1$). It is worthwhile to
imply here that the bound is exactly same as the one that has appeared in
ref.[39] where an application of the generalized distribution functions has
been discussed (see Fig.1 and Fig.3 of this reference).

Moreover, in the $x\equiv h\nu /kT<<1$ case, we verify that eq.(18) becomes

\begin{equation}
D_q(\nu)=\frac{8\pi (kT)^3}{h^2c^3}\frac{x^3}{x+\frac{(2-q)}{2!}x^2+
\frac{(2-q)(3-2q)}{3!}x^3+\;\cdots}\;\; ,
\end{equation}

\noindent which generalizes the Rayleigh-Jeans law. As it is expected, in
the $q\rightarrow 1$ limit this expression transforms to $D_1(\nu)\propto
x^2\propto\nu ^2$ which corresponds to the standard Rayleigh-Jeans law.

In addition to this, it is straightforward to generalize Stefan-Boltzmann
law and show that it remains the same, i.e. it is still proportional to
$T^4$, but with a $q$-dependent constant ($\sigma _q$). To see this, let us
write the total emitted power per unit surface,

\begin{equation}
P_q=\int_{0}^{\infty}d\nu D_q(\nu)\;\;\; .
\end{equation}

\noindent If we use here eq.(18) with the dimensionless variable $x$,

\begin{equation}
P_q=\frac{8\pi k^4}{h^3c^3}T^4 \int_{0}^{\infty} \frac{xdx}
{\left[1-(1-q)x\right]^{\frac{1}{q-1}}-1}
\end{equation}

\noindent can be obtained. Since the integral term is independent of $T$, it
can be written down as 

\begin{equation}
P_q=\sigma _q T^4 \;\;\; ,
\end{equation}

\noindent which is the generalized Stefan-Boltzmann law, with a
$q$-dependent prefactor.

It is also possible to obtain the generalized Wien shift law. By making
eq.(18) maximum, $\nu _m$ (the frequency value which makes $D_q(\nu)$
maximum) yields a nonlinear equation :

\begin{equation}
\left[\frac{h\nu _m}{kT}(3q-4)+3\right]\left[1-(1-q)\frac{h\nu _m}{kT}
\right]^{\frac{1}{q-1}}+\frac{h\nu _m}{kT}3(1-q)-3=0\;\; .
\end{equation}

\noindent Although it is very difficult to find an analytical solution for
this equation, the graphical solution is adequate for our purpose. From the
graphical solution, for $q=0.95$ and $q=1.05$ we have found $h\nu
_m/kT=2.444$ and $3.347$, respectively. For the same values of Tsallis
$q$-index, eq.(19) in ref.[24] gives $h\nu _m/kT=2.563$ and $3.08$. Once
again it is possible to see that our results provide bounds to the exact
ones. Moreover, the result $\nu _m(q>1)>\nu _m(q=1)>\nu _m(q<1)$, which
appears in ref.[24], is also valid in the present investigation. The
important diversity between this result and that of the $q_G$-deformed
quantum groups investigations namely $\nu _m(g_G)=\nu _m(1/q_G)$ and $\nu
_m(q_G<1)<\nu _m(q_G=1)$ is readily observed.

\section{Conclusions}

Although the generalized distribution functions are first introduced in
1993, there has been no attempt to apply them to the physical systems until
a very recent effort performed by Pennini et al. [39] where two single
particles are considered. In this letter, our goal is to apply one of the
generalized distribution functions (generalized Planck distribution) to
another physical system (Planck law for the blackbody radiation).
On the other hand, the approximate procedure used here to find a bound to
the generalized Planck law is simpler than that of Tsallis et al. [24]
(since the procedure is completely the same with that followed in standard
textbooks of Statistical Physics for the derivation of standard Planck law)
and also the bound seems to be more general, since not necessarily $q\cong
1$ (it will have a meaning, of course, if a physical system requiring this
condition exists).
Lastly, it is worthwhile to imply here that all the generalized laws derived
here (i.e. generalized Planck law, generalized Rayleigh-Jeans law,
Stefan-Boltzmann law and Wien law) transform to corresponding standard
well-known laws in the $q\rightarrow 1$ limit.

\section*{Acknowledgments}

The authors acknowledge TUBITAK and Ege University for making Prof. C.
Tsallis' visit to Izmir possible. We are very indebted to Prof. C. Tsallis
for the helpful discussions as well as kindly supplying to us some of the
references therein.

\newpage

\section*{Figure Captions}

\vspace{2cm}

\noindent
Blackbody photon energy density per unit volume versus $h\nu /kT$ in the
frame of this work and ref.[24].

\newpage

\begin{figure}[htbp]
\begin{center}
\addvspace{1cm}
\leavevmode
\epsffile{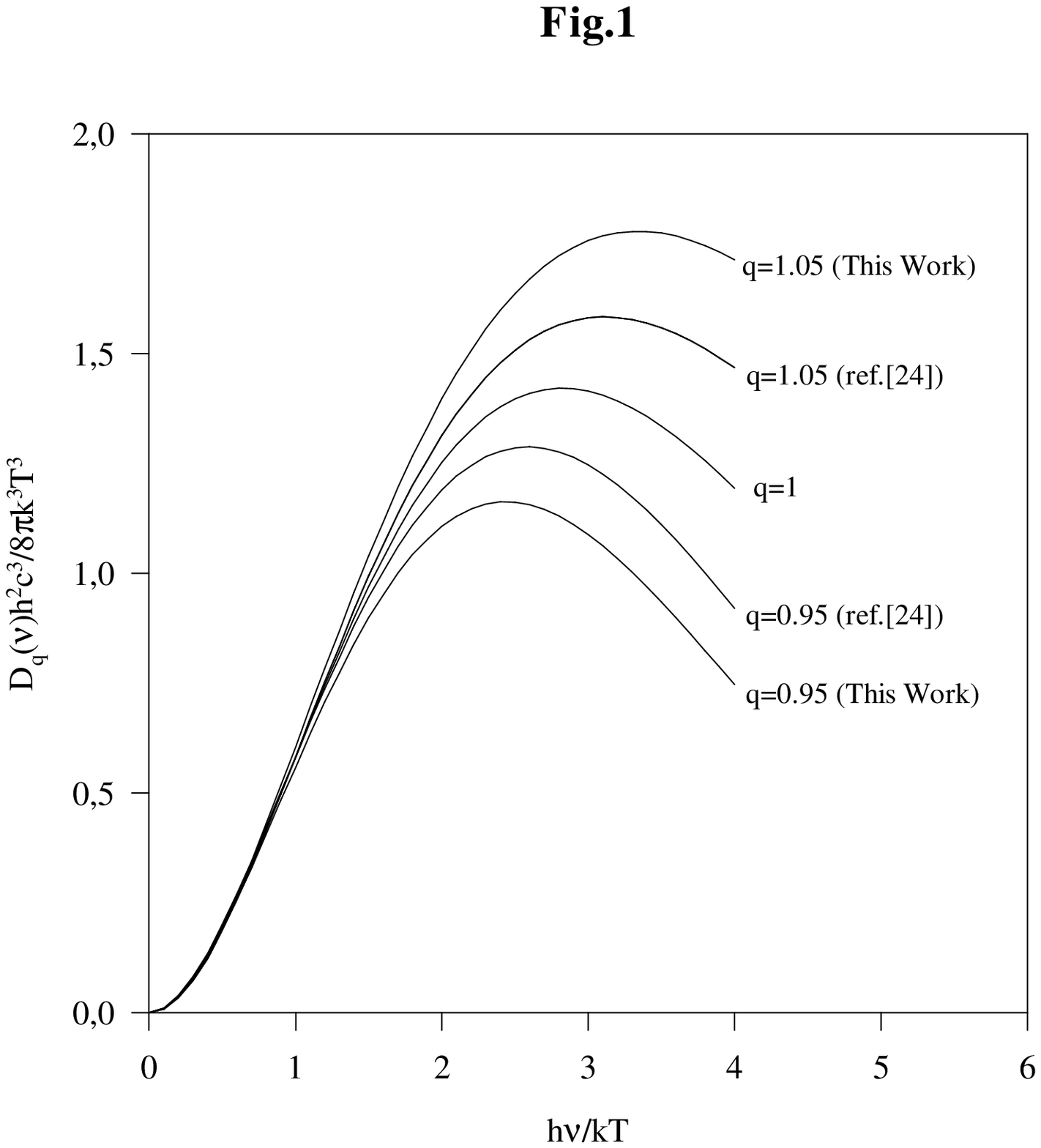}
\end{center}
\end{figure}

\end{document}